\documentclass[11pt,twoside]{article}
\usepackage{./asp2014}
\aspSuppressVolSlug
\resetcounters

\begin{document}
\title{Revealing the Large-Scale Structures of Interstellar Gas Associated with the Magellanic SNR N132D}
\author{H. Sano$^1$, Y. Fukui$^1$, S. Yoshiike$^1$, T. Fukuda$^1$, K. Tachihara$^1$, S. Inutsuka$^1$, A. Kawamura$^2$, K. Fujii$^2$, N. Mizuno$^2$, T. Inoue$^2$, T. Onishi$^3$, F. Acero$^4$, J. Vink$^5$
\affil{$^1$Department of Physics, Nagoya University, Furo-cho, Chikusa-ku, Nagoya 464-8601, Japan; \email{sano@a.phys.nagoya-u.ac.jp}}
\vspace*{-0.2cm}
\affil{$^2$National Astronomical Observatory of Japan, Mitaka 181-8588, Japan}
\vspace*{-0.2cm}
\affil{$^3$Department of Physical Science, Graduate School of Science, Osaka Prefecture University, 1-1 Gakuen-cho, Naka-ku, Sakai  599-8531, Japan}
\vspace*{-0.2cm}
\affil{$^4$Laboratoire AIM, CEA-IRFU/CNRS/Universit{\'e} Paris Diderot, Service d$'$Astrophysique, CEA Saclay, 91191 Gif sur Yvette, France}
\vspace*{-0.2cm}
\affil{$^5$Astronomical Institute “Anton Pannekoek”, University of Amsterdam, P.O. Box 94249, 1090 GE Amsterdam, The Netherlands
}}
\paperauthor{Hidetoshi Sano}{sano@a.phys.nagoya-u.ac.jp}{}{Nagoya University}{Department of Physics}{Nagoya}{Aichi}{464-8601}{Japan}
\paperauthor{Yasuo Fukui}{}{}{Nagoya University}{Department of Physics}{Nagoya}{Aichi}{464-8601}{Japan}
\paperauthor{Satoshi Yoshiike}{}{}{Nagoya University}{Department of Physics}{Nagoya}{Aichi}{464-8601}{Japan}
\paperauthor{Tatsuya Fukuda}{}{}{Nagoya University}{Department of Physics}{Nagoya}{Aichi}{464-8601}{Japan}
\paperauthor{Kengo Tachihara}{}{}{Nagoya University}{Department of Physics}{Nagoya}{Aichi}{464-8601}{Japan}
\paperauthor{Shuichiro Inutsuka}{}{}{Nagoya University}{Department of Physics}{Nagoya}{Aichi}{464-8601}{Japan}
\paperauthor{Akiko Kawamura}{}{}{National Astronomical Observatory of Japan}{}{Mitaka}{Kanagawa}{181-8588}{Japan}
\paperauthor{Kosuke Fujii}{}{}{National Astronomical Observatory of Japan}{}{Mitaka}{Kanagawa}{181-8588}{Japan}
\paperauthor{Norikazu Mizuno}{}{}{National Astronomical Observatory of Japan}{}{Mitaka}{Kanagawa}{181-8588}{Japan}
\paperauthor{Tsuyoshi Inoue}{}{}{National Astronomical Observatory of Japan}{}{Mitaka}{Kanagawa}{181-8588}{Japan}
\paperauthor{Toshikazu Onishi}{}{}{Osaka Prefecture University}{Department of Physical Science}{Sakai}{Osaka}{599-8531}{Japan}
\paperauthor{Fabio Acero}{}{}{CEA Saclay}{}{}{Gif sur Yvette}{91191}{France}
\paperauthor{Jacco Vink}{}{}{University of Amsterdam}{Astronomical Institute “Anton Pannekoek”}{}{Amsterdam}{94249}{The Netherlands}

\begin{abstract}
We report preliminary results of large-scale distribution toward the Magellanic supernova remnant N132D using Mopra and $Chandra$ archival datasets. We identified a cavity-like CO structure along the X-ray shell toward the southern half of it. The total mass of associating molecular gas is $\sim$$10^4 M_{\odot}$, which is smaller than the previous study by an order of magnitude. Further observations using ALMA, ASTE, and Mopra will reveal the detailed spatial structures and its physical conditions.
\end{abstract}

\vspace*{-0.5cm}
\section{Introduction}
\vspace*{-0.4cm}
In supernova remnants (SNRs), interaction between shock waves and surrounding interstellar gas is a key element for understanding the SNR evolution, cosmic ray acceleration, and origin of gamma-ray radiation. Recently, \cite{2015ApJ...799..175S} revealed that the shock-cloud interaction in a young Galactic SNR generates the turbulence and strong magnetic field, which enhance the non-thermal X-rays and an efficient acceleration of the cosmic ray electrons around the interacting gas clumps. Therefore, it is important to search for more evidence to examine its universality.\vspace*{0.2cm}

The SNR N132D is a bright X-ray emitter in the Large Magellanic Cloud (LMC; Figure \ref{fig1}a), and identified as a young SNR ($\sim$3150 yr). It is also known as an Oxygen-rich SNR \citep[e.g.,][]{2007ApJ...671L..45B} and is hence considered to be a remnant of a core-collapse supernova of a massive star. Furthermore, the SNR is thought to be possibly interacting with molecular gas as shown by CO observations with SEST \citep{1997ApJ...480..607B}. Therefore, it is the best target to study the shock-cloud interaction in the extra galaxies. In the present paper, we report the large-scale distribution of the associate molecular gas with N132D and its physical properties.

\articlefigure[width=0.9\textwidth]{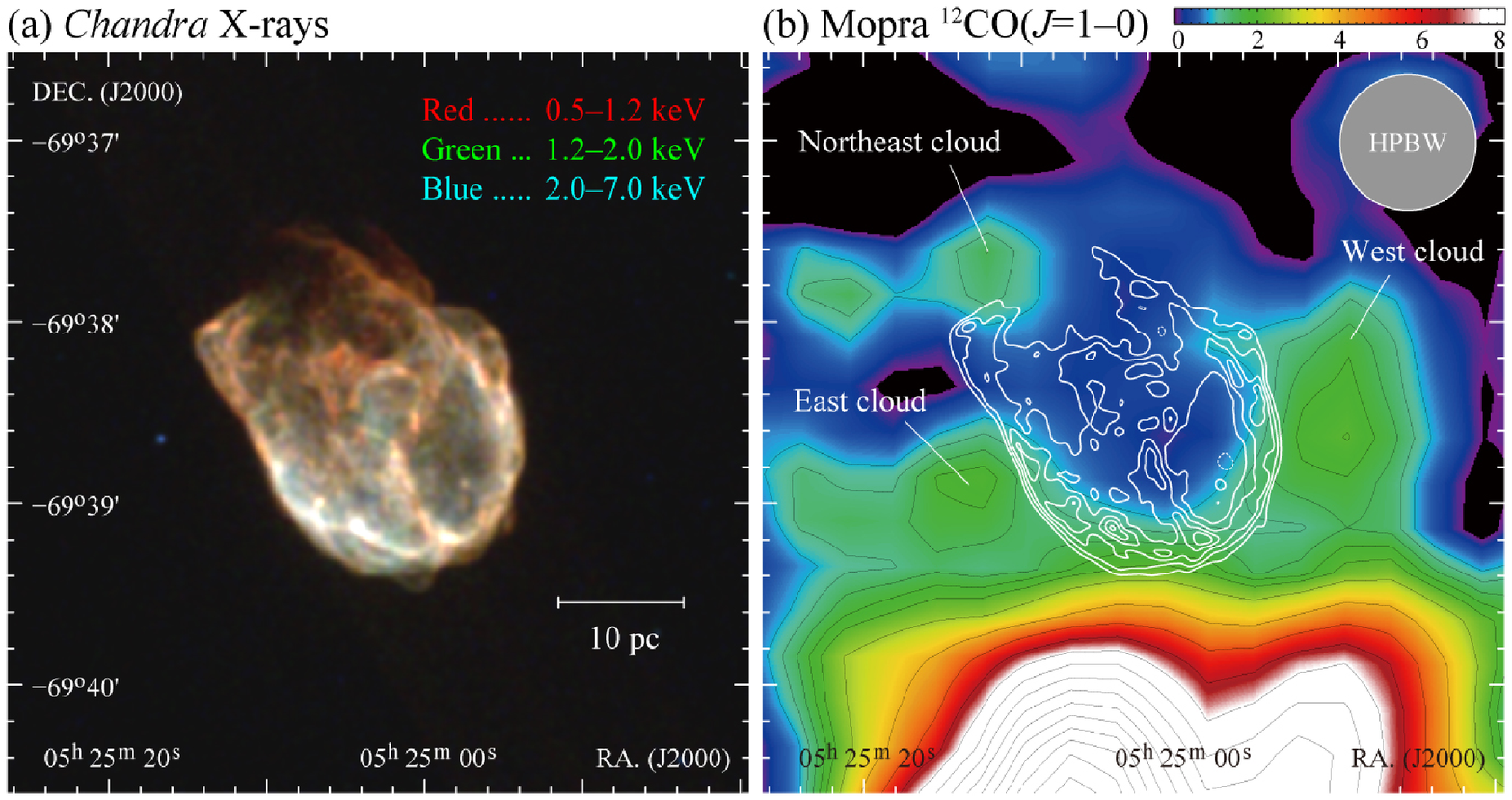}{fig1}{(a) The $Chandra$ X-ray tricolor image of SNR N132D \citep{2007ApJ...671L..45B}. (b) Integrated intensity map of $^{12}$CO($J$=1--0) \citep{2011ApJS..197...16W} in a velocity range of $V_{\mathrm{LSR}}$ = 256.5--268.6 km s$^{-1}$ is shown in color (unit is K km s$^{-1}$). The white contours correspond to the $Chandra$ X-ray flux in the energy band 0.5--7.0 keV.}

\vspace*{-0.35cm}
\section{Results}
\vspace*{-0.4cm}
Figure \ref{fig1}a shows $Chandra$ X-ray tricolor image of N132D \citep{2007ApJ...671L..45B}, which has shell-like and filamentary structures especially in the southwest, while the blowout structure appears in the northeast. Thermal X-rays (corresponding to a energy range of 0.5--1.2 keV) are bright over the whole SNR, whereas non-thermal X-rays (2.0--7.0 keV) emit only in the southern half.\vspace*{0.2cm}

Figure \ref{fig1}b represents the $^{12}$CO($J$=1--0) integrated intensity map taken by the Magellanic Mopra Assessment \citep[MAGMA;][]{2011ApJS..197...16W} with the Mopra 22-m telescope. The existence of the giant molecular cloud (GMC) in the southern part of Figure \ref{fig1}b was known by a previous study \citep{1997ApJ...480..607B}. We newly found three molecular clouds interacting with the SNR. The two clouds are located in the west and the east, which form the cavity-like CO structure along the X-ray shell. Another CO cloud is located in the the northeast. The total molecular mass of the newly found  3 clouds is $\sim$$10^4 M_{\odot}$ using the $X$-factor 7 $\times$ 10$^{20}$ [$W(^{12}$CO)/(K km s$^{-1}$)] (cm$^{-2}$) \citep{2008ApJS..178...56F}, which is smaller than the previous study by an order of magnitude \citep{1997ApJ...480..607B}. It is because the most part of the GMC is not interacting with the SNR.

\vspace*{-0.35cm}
\section{Discussion and Summary}
\vspace*{-0.4cm}
We discussed that the CO cavity-like structure was created by the stellar wind / UV photons from the massive star prior to the supernova explosion and is now interacting with the SNR shocks. Furthermore, the enhancement of non-thermal X-rays in the southern part can be described as a result of shock interaction with clumpy CO structures \citep[e.g.,][]{2015ApJ...799..175S}. Therefore, we predict that the surroundings of N132D have clumpy CO structures corresponding to the non-thermal X-ray filaments. We continue follow up observations with ALMA, ASTE, and Mopra, which allow us for the first time to study the interaction of a SNR with CO gas in detail outside our own Galaxy.

\end{document}